\documentclass[journal]{IEEEtran}

\usepackage[utf8]{inputenc}
\usepackage{acronym}
\usepackage{amsmath}
\usepackage{graphicx}
\usepackage{url}

\acrodef{CES}[CES]{Compton Event Space}

\title{Bounded-Latency Spherical-Histogram Reconstruction for Compton Cameras}

\author{Francisco~J.~Albiol,~Salvador~Tortajada,~Luis~Caballero,~José~Escalante,~Alberto~Albiol,~Elena~Larrea~Estrelles,~and~José~Luis~Leganés-Nieto%
\IEEEcompsocitemizethanks{
\IEEEcompsocthanksitem F.~J.~Albiol, S.~Tortajada, L.~Caballero, J.~Escalante, and E.~Larrea~Estrelles are with the Instituto de Física Corpuscular (IFIC), CSIC-Universitat de València, E-46980 Paterna, València, Spain. ORCID: F.~J.~Albiol 0000-0002-3824-2246; S.~Tortajada 0000-0001-7426-2407; L.~Caballero 0000-0002-1635-5282; J.~Escalante 0009-0008-9491-2938; E.~Larrea~Estrelles 0009-0002-0905-792X.
\IEEEcompsocthanksitem A.~Albiol is with the Campus de Vera, Universitat Politècnica de València, Camí de Vera s/n, 46022 Valencia, Spain. ORCID: 0000-0002-1970-3289.
\IEEEcompsocthanksitem J.~L.~Leganés-Nieto is with ENRESA, Spain. ORCID: 0000-0003-2901-6076.
}}

\begin{document}

\maketitle

\begin{abstract}
Gamma-ray imaging with Compton cameras remains computationally demanding because most analytical and iterative reconstruction pipelines preserve the list-mode nature of the acquisition inside the inversion loop. As the number of detected events grows, event-dependent cone/voxel interactions must be reevaluated repeatedly, limiting scalability and making low-latency reconstruction difficult. We present a spherical-histogram reconstruction framework in which each detected Compton event is encoded online into detector-centred angular histograms and volumetric reconstruction is then performed from coherent histogram snapshots through a sparse projection operator stored in memory. This change of representation turns the event stream into a bounded reconstruction state and decouples event accumulation from iterative inversion. In conceptual terms, the encoded spherical state acts as a computational hologram: a surface-encoded representation that preserves the angular information needed for volumetric inversion without replaying the photon list. The method supports multi-view and multi-resolution operation, non-blocking acquisition, and iterative forward/backward inversion whose dominant cost depends on the active sparse operator rather than on the total number of accumulated events. Dedicated timing measurements support the expected scaling split: conventional list-mode reconstruction time grows linearly with the number of events, whereas histogram-snapshot reconstruction remains approximately invariant with respect to accumulated event count once geometry, sparse operator, and iteration budget are fixed. Near-field Monte Carlo experiments demonstrate typically millimetric hotspot localization and bounded-latency reconstruction under continuous acquisition. The proposed formulation introduces a new reconstruction architecture for Compton imaging while leaving detector-specific calibration refinements and broader application studies for future work.
\end{abstract}


\section{Introduction}

\IEEEPARstart{G}{amma}-ray imaging plays a crucial role in medical diagnostics, image-guided interventions, and treatment monitoring~\cite{Kim2024, Parajuli2022}. Unlike visible light, gamma radiation cannot be focused using conventional lenses or mirrors, making the determination of its origin a fundamentally challenging problem. Among the different physical processes governing the interaction of gamma rays with matter, Compton scattering provides a unique opportunity to infer directional information without the need for mechanical collimation.

Compton scattering occurs when a gamma-ray photon collides with an electron, transferring part of its energy to the electron and being deflected from its original path. The amount of energy exchanged during this interaction is directly related to the scattering angle of the photon. As a consequence, a single interaction does not reveal the exact incoming direction of the radiation, but constrains it to lie on the surface of a cone in three-dimensional space. This geometrical property forms the physical foundation of Compton imaging.

A Compton camera exploits this principle by measuring both the energy and the position of sequential photon interactions within a detector system. Each detected event defines a so-called {\em Compton cone}, representing all possible directions from which the photon could have originated. While a single event provides only a cone of uncertainty, the accumulation of many such events allows these cones to overlap and converge toward the true source location. The reconstruction of the gamma-ray source therefore becomes a problem of identifying the region in space where a large number of cones intersect.

In practice, this reconstruction can be formulated in terms of angular projections derived from the detected photon trajectories. Line projections are established based on the measured direction of scattered photons, while Compton cones encode the set of admissible trajectories compatible with the energy deposits in the detector. Image formation then relies on analyzing the intersections and overlaps of these projections and cones to determine the most probable source location in three-dimensional space.

Despite its clear geometrical interpretation, the reconstruction process is computationally demanding. It requires processing large numbers of photon events and combining their associated constraints using advanced statistical and data-processing techniques to obtain an accurate image of the gamma-ray source. These problems are commonly addressed using a list-mode approach, in which each detected event is treated individually and its cone contribution is preserved inside the reconstruction loop.

These techniques include list-mode processing combined with Maximum Likelihood Expectation-Maximization, which achieves accurate results but demands very high computational time~\cite{Barrett1997, Wilderman1998, Wilderman2001, Tornga2009, Kolstein2014, Le2023}; the Filtered Back Projection analytical reconstruction method, which is faster but suffers from significant inaccuracies in noisy data~\cite{Parra2000, Tomitani2002, Hirasawa2003, Xu2006, Truong2007}; the Bayesian Maximum A Posteriori approach, which incorporates regularization to slightly reduce noise~\cite{Sauve1999, DePierro2001, Ahn2003, Lee2008}; and Markov Chain Monte Carlo methods, which offer linear reductions in reconstruction time~\cite{Andreyev2011, Andreyev2016}. More recently, deep learning-based approaches have shown promise, delivering slightly better reconstruction quality and reduced processing times compared to traditional methods~\cite{Ikeda2021, Daniel2022}; in prompt-gamma range imaging for proton therapy, machine-learning-aided Compton approaches have also been proposed specifically to address the need for real-time operation under high count-rate and high-background conditions~\cite{LerendeguiMarco2022}. Even when GPU acceleration is employed~\cite{Nguyen2016}, however, the core difficulty remains event dependent: the inversion still has to revisit cone-specific weights that depend on scattering angle, detector geometry, attenuation, voxel intersection length, or some approximation thereof. As a consequence, the cost of reconstruction typically grows with the number of accumulated events and it becomes difficult to maintain bounded latency when the acquisition is continuous.

The limitation of current Compton imaging systems is therefore not only one of image quality, but also one of representation. List-mode capture preserves the raw event stream, but it does not provide a compact reconstruction state that can be used efficiently under continuous acquisition.

\begin{figure*}[t]
    \centering
    \includegraphics[width=\textwidth]{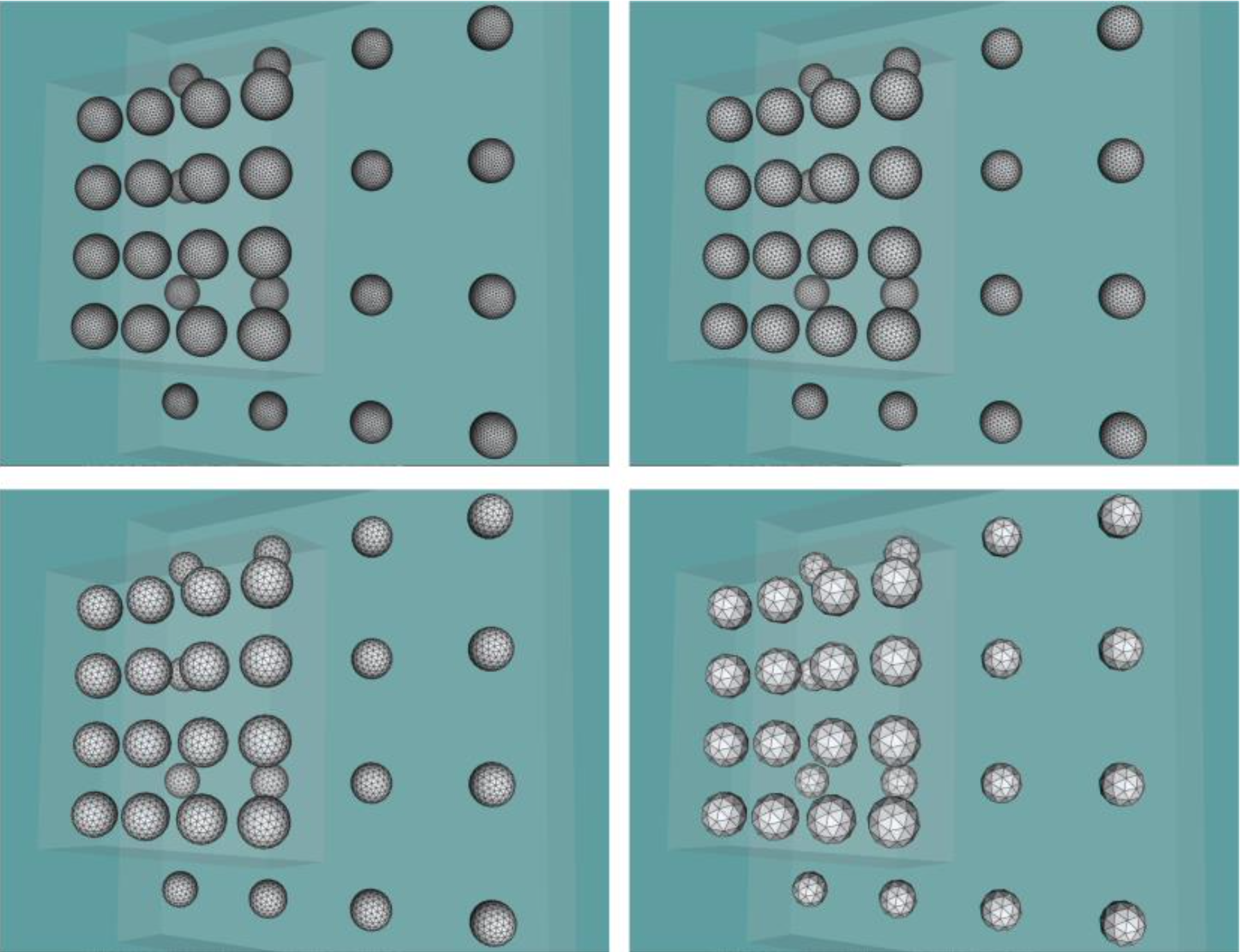}
    \caption{Illustration of the virtual decomposition of a Compton camera to represent the Compton Event Space with spherical histograms. This particular example uses a front detector and a parallel larger back detector. Four different configurations with different resolutions are shown.}
    \label{fig:spherical_histograms}
\end{figure*}

Instead of retaining event geometry exclusively in list form, we propose to encode each Compton cone into detector-centred spherical histograms arranged as a fly-eye structure (Fig.~\ref{fig:spherical_histograms}). The intersection of each cone with a sphere becomes a circular support on a geodesic angular domain, and each spherical bin acts as a virtual pixel defining a candidate ray. Several spherical histograms can then be combined to recover a three-dimensional activity volume.

The key contribution of this paper is the transition from event-wise list processing to a spherical-histogram representation of the Compton Event Space. The event stream is first encoded into bounded angular states; reconstruction is then performed from coherent snapshots through a sparse operator defined in memory. This allows acquisition and inversion to be decoupled and makes the dominant cost of iterative reconstruction depend on the sparse model rather than on the full history of events. More specifically, the \ac{CES} should be understood as a localized spatial-angular representation of the Compton process: the spatial arrangement of the fly-eye spheres supplies local detector support, while the spherical bins attached to each sphere supply local angular support. This differs from a purely global angular description, because spatial locality, directional locality, detector geometry, and multi-view parallax remain explicit in the encoded state. In this sense, the bounded spherical state can be interpreted as a computational hologram: the source field is not stored as a replayable photon list, but as a distributed coded angular surface from which the activity volume can be reconstructed. The paper establishes this representational and algorithmic layer as the foundation on which detector-specific calibration, self-calibration, and image-quality optimization can be composed.

\section{Materials and Methods}

\subsection{Compton Camera\label{ssec:mm_compton_camera}}

A Compton camera is a device for detecting the direction of incoming gamma-ray photons. In its simplest form, it consists of two detectors arranged one behind the other in parallel. Based on the principles of Compton scattering, incident gamma-rays interact with a scatterer detector, resulting in scattered photons and recoil electrons. The scattered photons undergo further interaction with the absorber detector. Both detectors generate signals that encode information about the position of the interaction and the deposited energy. Using this information we can infer the direction of the incident gamma-ray photons and locate their sources in three-dimensional space by means of a Compton cone. The surface of a Compton cone represents the possible paths of origin from the source. The direction of the cone can be inferred from the straight line formed by the positions of the interactions with the detectors. The angle of the cone is defined by the energies deposited on each detector according to the relation described in~(\ref{eq:compton_angle}).

\begin{equation}
    \cos{\theta} = 1-\frac{m_e c^2 E_r}{E_s(E_r + E_s)}
    \label{eq:compton_angle}
\end{equation}

\noindent where $\theta$ is the Compton angle, $m_e$ is the electron mass, $c$ is the speed of light, $E_r$ is the recoil electron energy detected by the scatterer and $E_s$ is the energy of the scattered photon detected by the absorber detector.

\begin{figure*}[t]
    \centering
    \includegraphics[width=\textwidth]{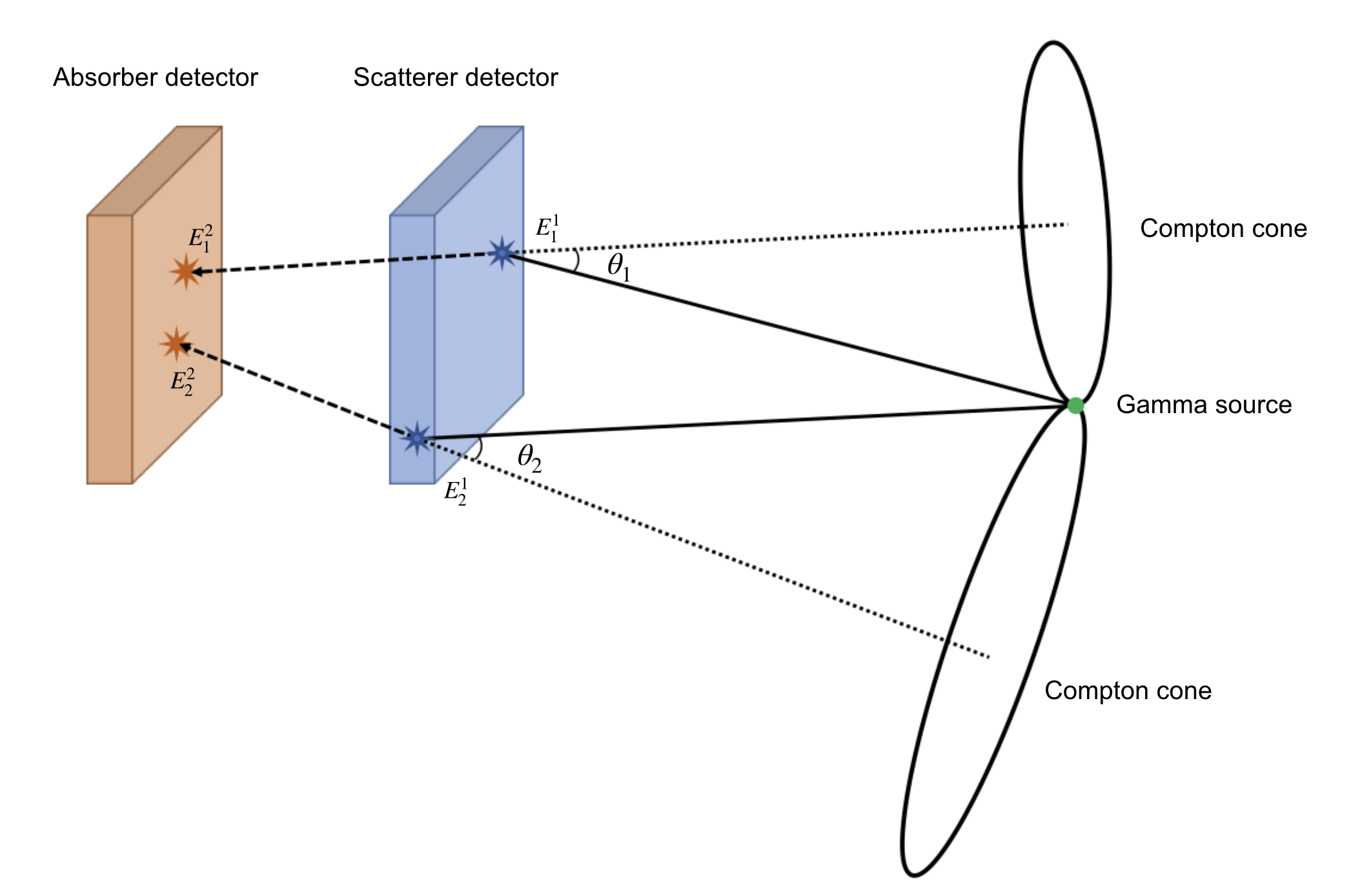}
    \caption{Two photons, coming from the same source, produce two hits in the scatter detector plane. The Compton cone containing the original source can be recovered by the information in the absorber detector after the Compton event occurs.}
    \label{fig:compton_events}
\end{figure*}

When a set of Compton events are detected, they each describe a Compton cone, which makes it possible to find the position of the gamma-ray source at the location where the cones intersect most frequently (see Fig.~\ref{fig:compton_events}).

\subsubsection{Detector Setup and Data Acquisition\label{ssec:mm_detector_setup}}

\begin{figure*}[t]
    \centering
    \includegraphics[height=0.28\textheight]{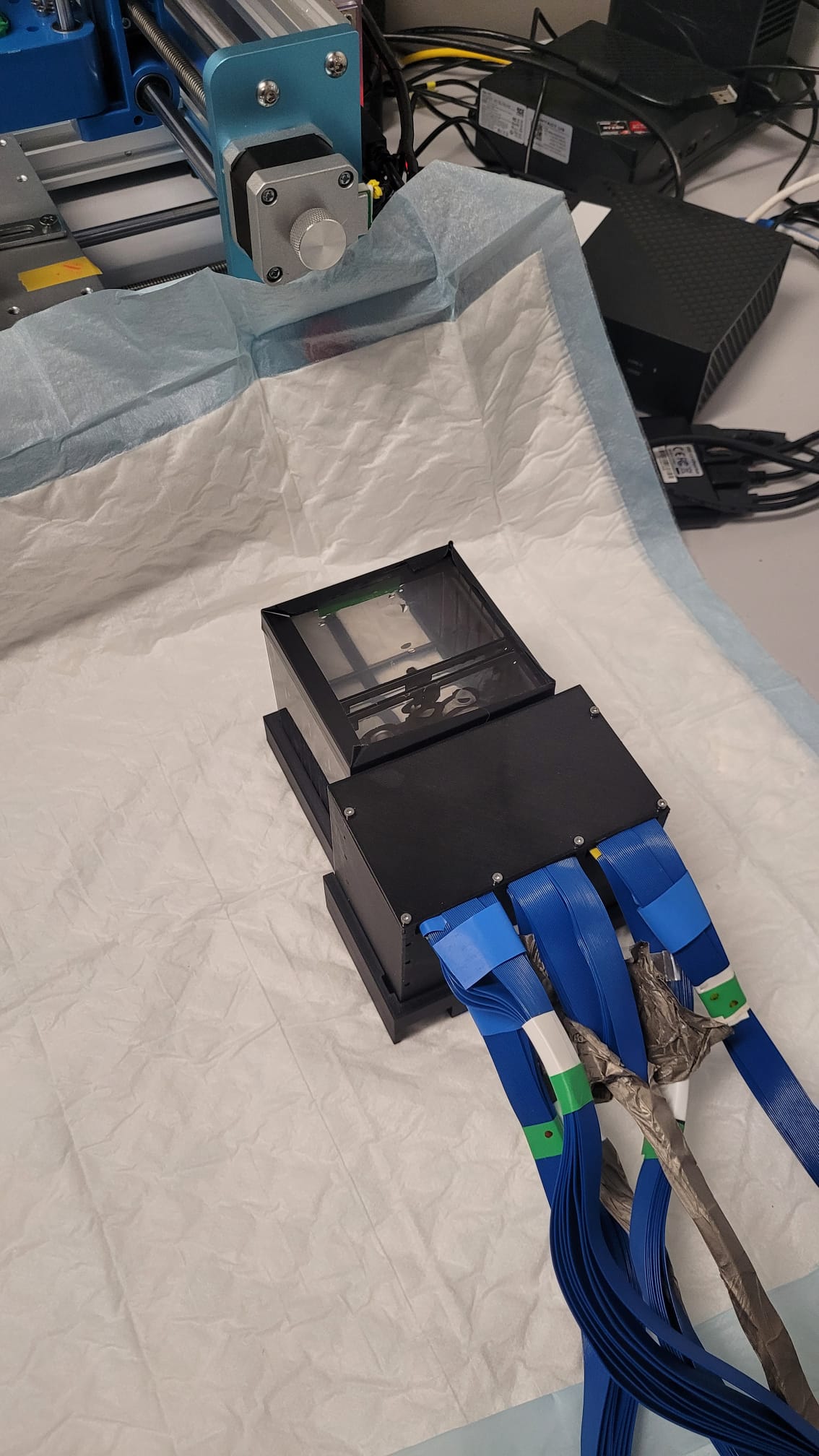}\hfill
    \includegraphics[height=0.28\textheight]{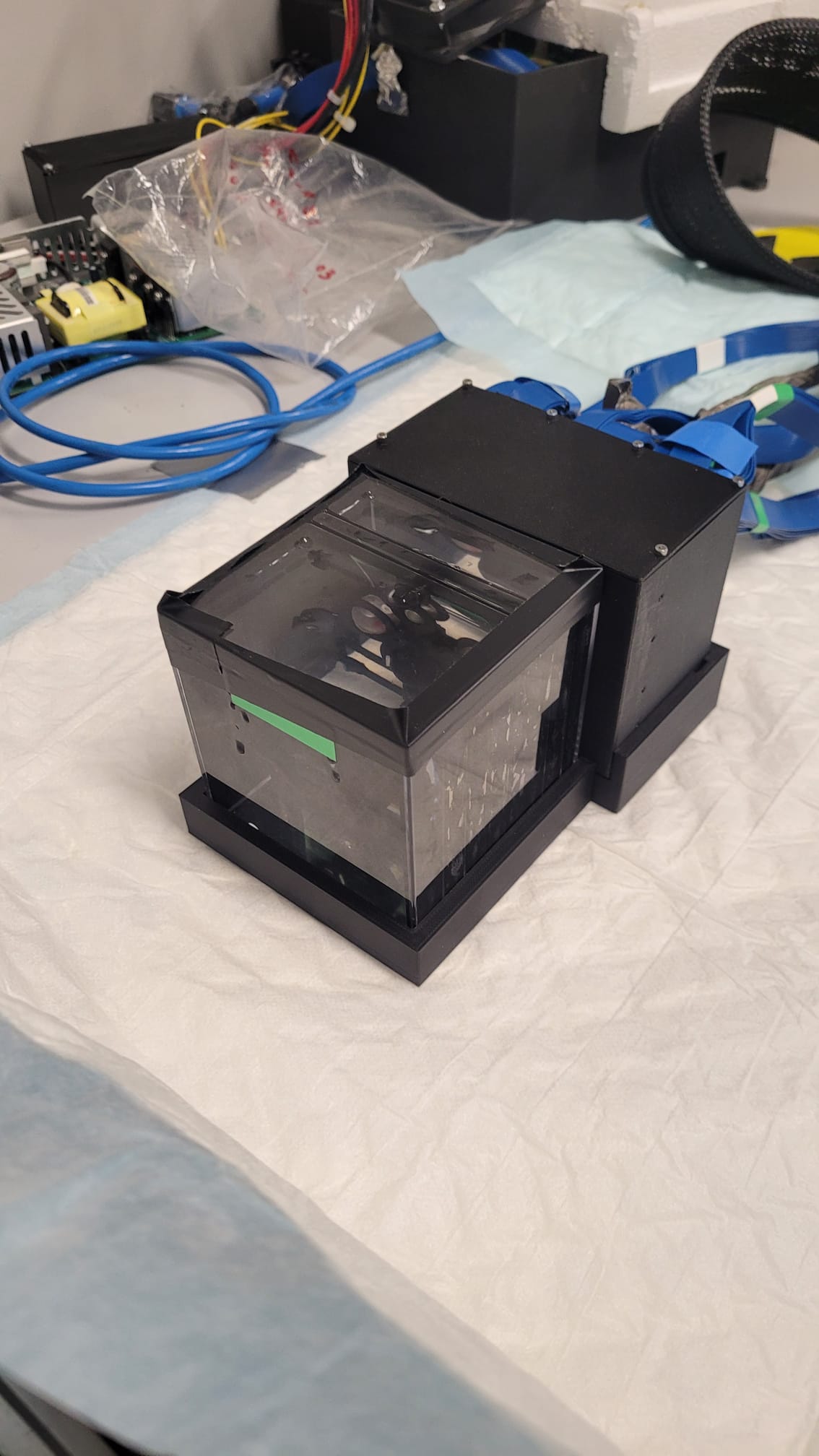}\hfill
    \includegraphics[height=0.28\textheight]{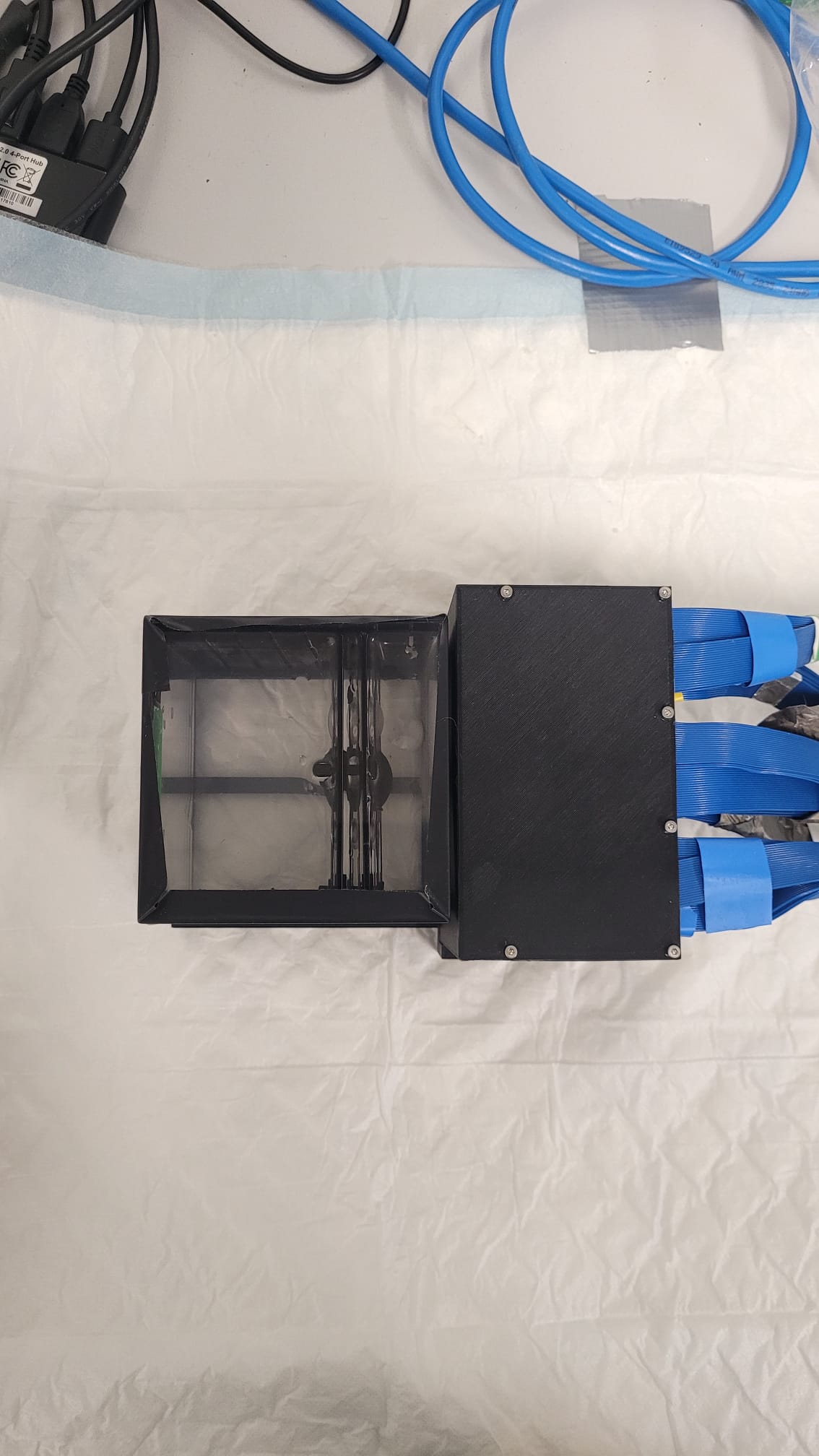}
    \caption{Experimental measurement setup. The photographs show the PETsys-connected detector head, the source-positioning stage, and the shielding/background arrangement used for the phantom measurements. The six-crystal scatter--absorber geometry used by the event builder is specified by the instrumental configuration; PETsys provides the acquisition electronics used to form calibrated Compton candidates.}
    \label{fig:experimental_setup}
\end{figure*}

The experimental camera is a six-crystal GAGG Compton detector (Fig.~\ref{fig:experimental_setup}). Three $25.5 \times 51.0 \times 5.0$~mm$^3$ crystals form the scatter layer and three $25.5 \times 51.0 \times 8.0$~mm$^3$ crystals form the absorber layer. In each layer, the module centres are located at $x=\{-32.9,\,0,\,32.9\}$~mm. The scatter and absorber centres lie at $z=-2.5$~mm and $z=-33.0$~mm, respectively, giving a 24-mm inter-plane gap. This configured geometry is used by the event builder to associate the two calibrated interactions forming a Compton candidate with their physical crystal modules.

Signals are acquired with conventional PETsys readout electronics; the blue ribbon cables visible in Fig.~\ref{fig:experimental_setup} connect the detector head to that commercial acquisition chain. The PETsys configuration supplies the bias and discriminator settings, TDC and QDC calibration tables, and channel and trigger maps. The work reported here does not introduce new front-end electronics. Its instrumental role is limited to converting the calibrated PETsys stream into Compton candidates, using the channel map and the known detector geometry.

\subsubsection{Preparation of Compton Events\label{ssec:mm_commissioning}}

The PETsys output is first associated with the channel-to-pixel and pixel-to-crystal map. Pixel hits are grouped in a 10,000-ps window centred on the reference timestamp, and the resulting calibrated singles provide deposited energy, timestamp, detector identifier, and interaction position. A second 10,000-ps centred window groups singles from different crystal modules into candidate scatter--absorber interactions; their ordering is assigned from the two calibrated energy deposits. The event builder applies the acquisition-specific energy selection before emitting a Compton event. Thus, the PETsys calibration tables, channel map, pixel and event time windows, energy selection, and six-module geometry are the instrumental inputs to the Compton-event stream; the spherical-histogram reconstruction starts only after this event definition.

The evaluation combines measured and simulated data, which are kept distinct throughout the analysis. The structured-phantom measurements use $^{18}$F-FDG activity and therefore 511-keV annihilation photons acquired with the PETsys chain described above; the acquisition used for the reported phantom example lasted 5~min. Candidate events were selected with a 4\% energy criterion relative to the 511-keV photopeak (about 20~keV), consistent with the energy resolution measured with a $^{22}$Na reference source and observed with the FDG data. This selection criterion should not be confused with the uncertainty of the Compton energy sum: because the two deposits lie below the photopeak, the propagated relative uncertainty of their sum is approximately 6\%. The controlled near-field localization study is a Geant4 Monte Carlo experiment at 662~keV, used to isolate geometrical localization and computational-scaling behaviour. Consequently, the Monte Carlo results do not constitute a replacement for calibration or response characterization of the measured PETsys system.

\subsection{Spherical Histograms as Cameras\label{ssec:mm_spherical_histograms}}

In camera projection, the usual pinhole model~\cite{Hartley2000} uses a ray that projects a point in three-dimensional space to the centre of projection. This ray intersects a two-dimensional plane which is the image plane. The intersection point in a digital camera is contained in a pixel, which records the intensity of the captured light. Conversely, the pixel describes a ray of the two-dimensional image plane in the direction of the original three-dimensional point.

In contrast, neither the hit pixel of the scatterer or absorber in Compton cameras represents a single ray, and the combination represents a cone which itself has computational issues to be represented or projected in a voxelized/pixelized space.

Our proposal is to make a discretization of the surface of scatterer and absorber as a set of different spherical surface histograms. Geometrical surface histograms representations are a natural representation of a cone with the origin in the center of the sphere, which guarantees that the intersection between the sphere and the cone defines a circumference.

The treatment of each sphere as a spherical histogram is analogous to the treatment of the two-dimensional image plane, where the intensity is collected in each pixel of the image histogram. Likewise, the light intensity is collected in each of the bins of our spherical histogram by probable photon-counting. In this way, each bin, like a virtual pixel, projects a ray in the corresponding direction, marking the way in which the original source can be located. Unlike a conventional camera, however, the spherical histogram is also a persistent state variable: it accumulates the event stream online and can be frozen, stored, compared with past states, or reconstructed into a volume without replaying the full list of raw events.

To carry out this approximation, we first decompose the detectors into a set of fly-eye-like structure using spheres. Then, we project the Compton cone on the sphere. This sphere is divided into a number of virtual pixels or bins that allow each cone to be represented as a circular histogram. As usual, the number of bins defines the resolution.

\subsection{Virtual Decomposition of the Detector\label{ssec:virtual_decomposition}}

The aim of this approach is to construct a hyper-surface that accurately captures the Compton interactions within the detector, ensuring that the sampling process adheres to the principles of the Sampling Theorem, thereby avoiding as much as possible aliasing and loss of crucial information. The spherical histograms are a virtual representation of the detectors, where each detector is decomposed into an arbitrary number of spheres that cover its entire space. These spheres represent any Compton cone when the detector acts as a scatterer. The spheres are like a fly-eye structure. It is important to distinguish between the real pixels of the detector and the fly-eye representation created by the spheres. Thus, the fly-eye corresponds to the Compton interactions detected by the real pixels covered by the corresponding sphere. Each fly-eye represents more than one real pixel of the detector, which means that there is a loss of spatial resolution that depends on the configuration. The decision to use a larger or smaller number of spheres to decompose the detector may depend on the available statistics, the angular resolution and the angular error.

\begin{figure*}[t]
    \centering
    \includegraphics[trim={14cm 8cm 14cm 14cm}, width=\textwidth]{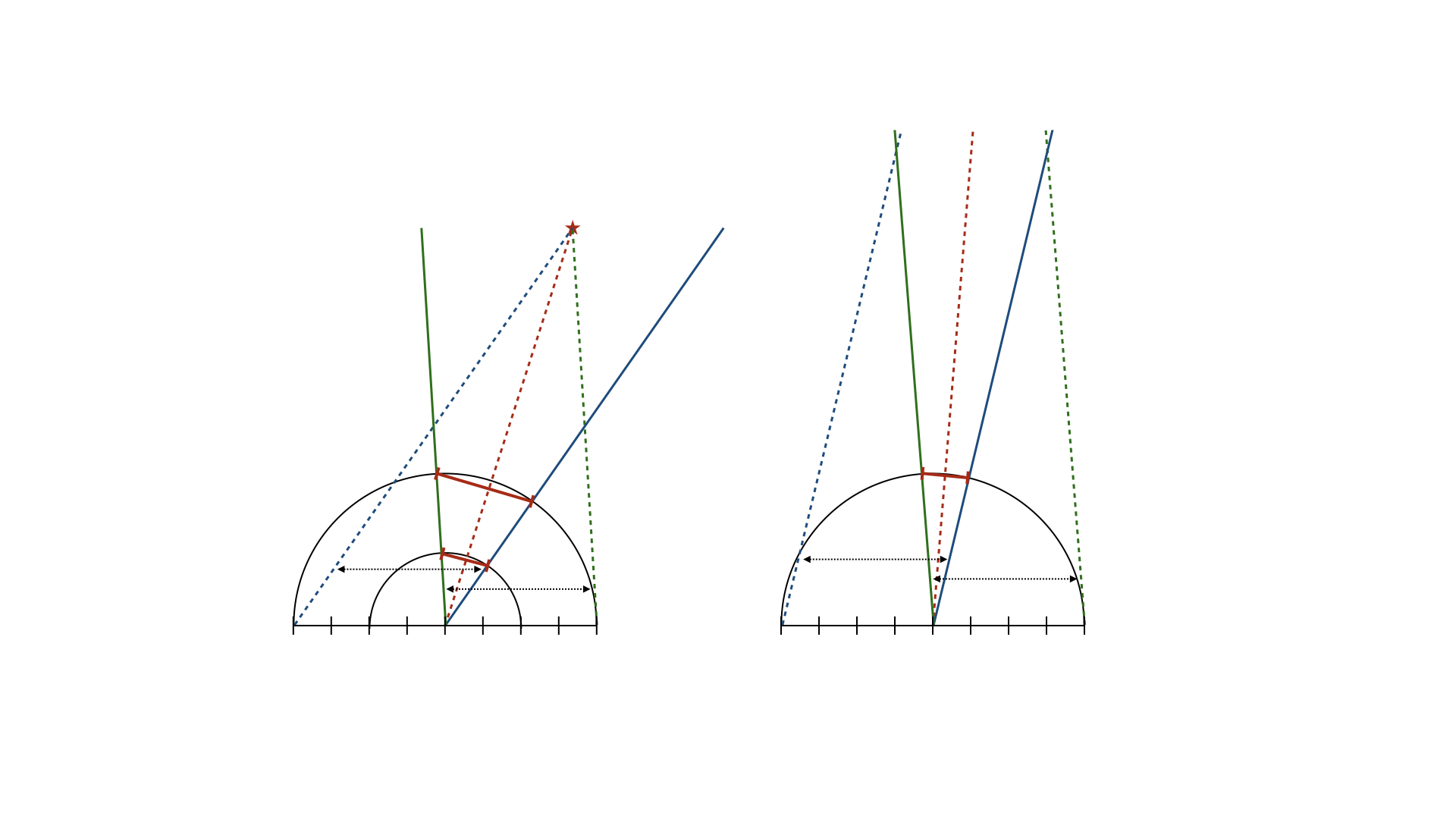}
    \caption{An illustration of the angular error (red segment) for different fly-eye structures and sources at different distances. On the left, we can compare the angular error for fly-eyes where one has a diameter twice as big as the other. On the right, we can compare the angular error when the source is farther than the one in the left image. It can be seen that the angular error depends on the size of the fly-eye and on the distance of the source to the detector.}
    \label{fig:virtual_pixel}
\end{figure*}

The angular error is determined by the size of the fly-eye, the distance from the source to the detector and the angle of the source with respect to the plane of the detector. Any real pixel is represented by its corresponding fly-eye structure by shifting it to its centre, as explained in Section~\ref{ssec:mm_projection_compton_cones}. Therefore, it can be seen that the error depends on the angle formed by the edges of the fly-eye when moved to the centre and that this angle depends in turn on the distance from the source to the detector (see Fig.~\ref{fig:virtual_pixel}).

In the limit case where the source is at infinity, the fly-eye approximation error tends to zero. Even in that regime, however, using several fly-eyes instead of a single global one remains advantageous because the angular response of each view creates structured directional patterns that can later be exploited during volumetric reconstruction.

\subsection{Projection of Compton Cones and Bin Counting\label{ssec:mm_projection_compton_cones}}

\begin{figure*}[t]
    \centering
    \includegraphics[width=\textwidth]{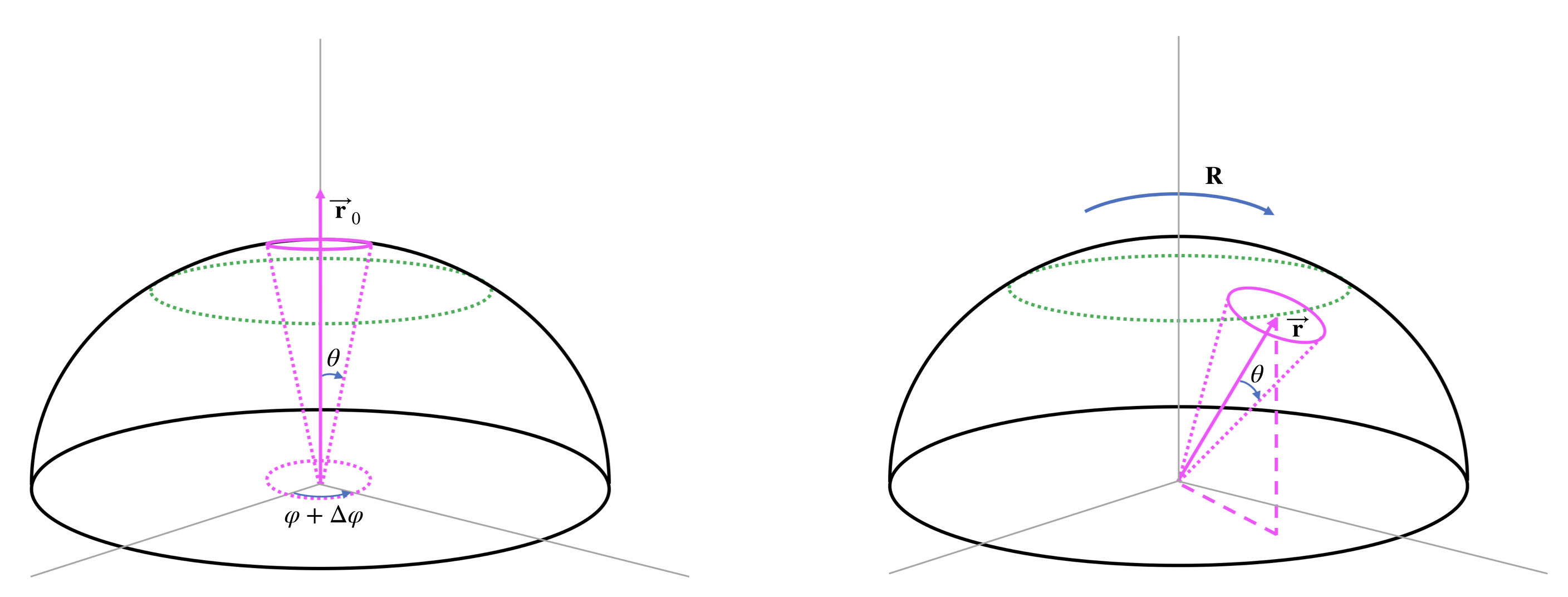}
    \caption{Schematic illustration of the computation of the intersecting points $\mathcal{P}$ defined by the intersection between the sphere and the Compton cone with the direction and angle estimated from the scatterer and absorber information. The hemisphere is shown, but the rotation is applied to the whole sphere.}
    \label{fig:cone_sphere}
\end{figure*}

The projection of each Compton cone depends on the information provided by the detected positions and the energies recorded in the scatterer and absorber detectors. The projection follows these steps:

\begin{enumerate}
    \item Take an initial direction pointing at the zenith of the sphere $\vec{\mathbf{r}}_0$.
    \item The Compton angle is estimated using the energies detected by the scatterer and absorber (see~(\ref{eq:compton_angle})). The Compton angle sets the polar angle, $\theta$, that defines the Compton cone intersecting the sphere with direction $\vec{\mathbf{r}}_0$.
    \item The intersecting points, $\mathcal{P}$, between the Compton cone and the surface of the sphere are calculated using an azimuthal angle, $\varphi$, varying from $0$ to $2\pi$ with steps depending on the resolution, i.e. the number of virtual pixels ($N$): $$\Delta \varphi=\frac{2\pi}{N-1}$$
    These first steps are illustrated in Fig.~\ref{fig:cone_sphere}, left.
    \item The final direction $\vec{\mathbf{r}}$ is estimated from the detected positions in the scatterer and absorber detectors.
    \item A rotation $\mathbf{R}$ is calculated from the initial and final directions. This rotates the sphere by superimposing the direction of the zenith with the calculated direction of the Compton event $\vec{\mathbf{r}}$.
    \item The intersecting points $\mathcal{P}$ are also rotated with $\mathbf{R}$.
    \item Finally, the bins through which the described circumference passes, i.e. points $\mathcal{P}$, are calculated and added to the counts of these bins as a spherical histogram (Fig.~\ref{fig:cone_sphere}, right).
\end{enumerate}

For a given sphere $s$, channel $\ell$, and bin $b$, the online event update can be written as
\begin{equation}
H_{s,b}^{(\ell)} \leftarrow H_{s,b}^{(\ell)} + \alpha_{e,b},
\end{equation}
where $\alpha_{e,b}$ is the anti-aliased contribution of event $e$ to bin $b$. In practice, each point of $\mathcal{P}$ contributes to one or more bins and the per-cone weights are normalized so that
\begin{equation}
\sum_{b \in \mathcal{C}(e)} \alpha_{e,b} = 1,
\end{equation}
with $\mathcal{C}(e)$ denoting the discrete circular support of the cone on the sphere. This normalization preserves a consistent event mass when the same cone is represented at different angular resolutions. To avoid aliasing, oversampling and grouped bin updates can be used.

The main advantage of this cone-update stage is computational cost rather than a claim of maximal geometric accuracy. Instead of exhaustively sampling the full spherical surface, or equivalently traversing a dense volumetric grid with pointwise angular tests, the proposed procedure restricts the computation to the discrete support of the cone on the selected sphere. The anti-aliased update therefore concentrates work on the geometrically relevant locus of the event and avoids the cost of uniform dense evaluation. This cost-oriented formulation is what makes the method suitable for sustained online operation when many events must be integrated continuously.

As mentioned, each bin of the spherical histogram represents a ray. These rays indicate a possible direction in which to locate the original source. As in traditional histograms, the probability of each ray will be higher the higher the count number. It should be stressed that, unlike traditional histograms, the counts of each bin do not represent a detected photon but one of the possible angular support locations compatible with the Compton event described by the set of points $\mathcal{P}$.

\subsection{Projection of Rays from the Spherical Histogram\label{ssec:mm_projection_rays}}

The bins of a spherical histogram are determined in spherical coordinates. However, to reconstruct the three-dimensional source location we need Cartesian coordinates. Assuming spheres with unit radius ($\rho=1$), the spherical coordinates of each bin can be converted to Cartesian coordinates directly by means of the following well-known transformations:
\begin{equation}
\begin{split}
r_x &= \sin{\theta}\cos{\varphi}\\
r_y &= \sin{\theta}\sin{\varphi}\\
r_z &= \cos{\theta}\\
\end{split}
\end{equation}
\noindent where $\theta$ is the polar angle and $\varphi$ is the azimuthal angle of a bin (or virtual pixel) in the sphere~\cite{ISO80000}\footnote{We use the physics convention of the spherical coordinates defined by the ISO 80000-2.}.

Thus, each bin projects a ray defined by the director vector in Cartesian coordinates. Each ray points to a direction and the most probable direction is the one that belongs to the bin with the highest frequency in number of counts.

\subsection{Efficient Processing with Multi-Resolution}

Multi-resolution allows to analyze or represent data at multiple levels of detail or resolution. In this work, it is employed to efficiently handle data by representing it in a hierarchical structure. This involves representing the Compton cones at different levels of detail, from coarse to fine, allowing for efficient storage and processing. Furthermore, using multi-resolution techniques enables resolution adjustments to be executed at a unit computational cost, thereby facilitating the selection of the most suitable resolution based on the signal-to-noise ratio. The multi-resolution setting is carried out by means of different layers of resolution that can be selected manually or automatically.

\begin{figure*}[t]
    \centering
    \includegraphics[width=\textwidth]{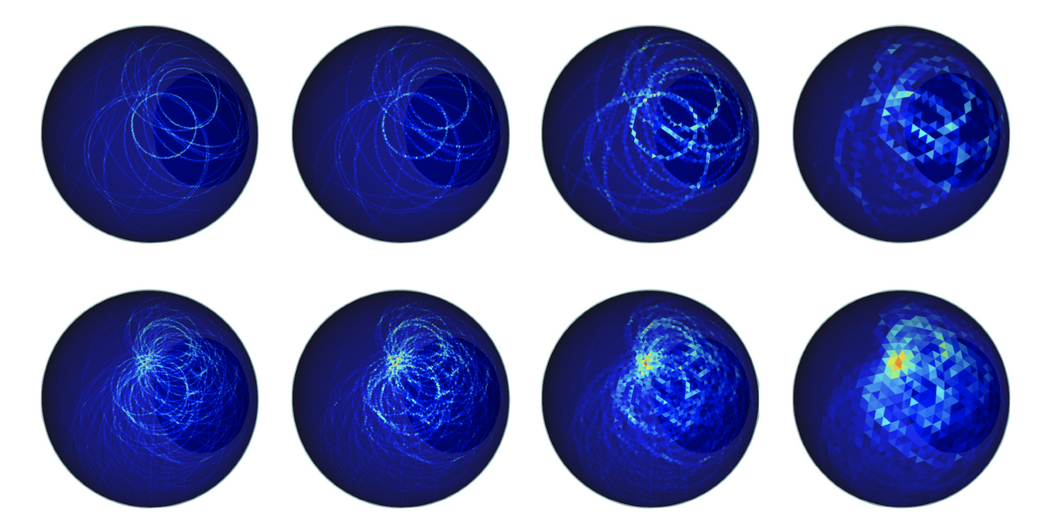}
    \caption{Spherical histograms with different resolutions. All these resolutions are stored in memory, hence any resolution configuration is easily accessible.}
    \label{fig:multi-resolution}
\end{figure*}

\subsection{Configuration of a Multi-View Camera\label{ssec::mm_multiview}}

A multi-view camera is a system that captures images from multiple angles simultaneously to provide a three-dimensional view of an object. As mentioned in Section~\ref{ssec:mm_spherical_histograms}, the arrangement of several spheres in the scatterer and absorber planes allows to make a discretization of their surface turning our approach into a multi-view camera-like configuration. This is the so-called \ac{CES} representation.

When detecting multiple Compton events, statistics are accumulated. Hence, the intersections of the Compton cones with the sphere allow the accumulation of counts in their bins, as explained in Section~\ref{ssec:mm_projection_compton_cones}. These bins work as virtual pixels whose direction is a ray pointing towards the source. Thus, the most likely bins are those where more Compton cones intersect each other.

Furthermore, the different spherical histograms of each sphere project the most probable rays into Cartesian coordinate space. This makes it possible to locate the most probable region in which the emitting source is located by crossing the rays of each spherical histogram in a similar way as one would do in a multi-view camera with flat images.

The configuration of the \ac{CES} as a multi-view camera allows to capture multiple perspectives simultaneously, which solves the parallax limitation of a single Compton camera and enables the reconstruction of three-dimensional objects.

This multi-view interpretation is also useful for understanding why the representation is more than a compact histogram. Each sphere acts as a localized directional observer of the same underlying activity volume. A true source therefore produces angular patterns that are shifted and transformed from sphere to sphere, but remain geometrically correlated across the fly-eye arrangement. Random coincidences, statistical fluctuations, and geometrically inconsistent scatter are less likely to reproduce the same multi-view consistency. The useful signal can therefore be interpreted as occupying a coherent spatial-angular structure in \ac{CES}, while a substantial fraction of noise remains less coherent across views.

From a computational point of view, this representation also decomposes the online update stage into a collection of independent state objects, one per sphere. Each object stores its own spherical state and can therefore be updated through an independent producer--consumer workflow. This object-level decomposition provides a natural form of parallelization: the global acquisition stream is absorbed by multiple sphere-local update units rather than by a single monolithic accumulator. Load is not expected to be perfectly uniform across spheres, so the sustainable operating point is determined by the most demanding sphere-local queue. Even so, the decomposition is valuable because it localizes contention and allows the remaining spheres to continue integrating events concurrently.

\begin{figure*}[t]
    \centering
    \includegraphics[width=\textwidth]{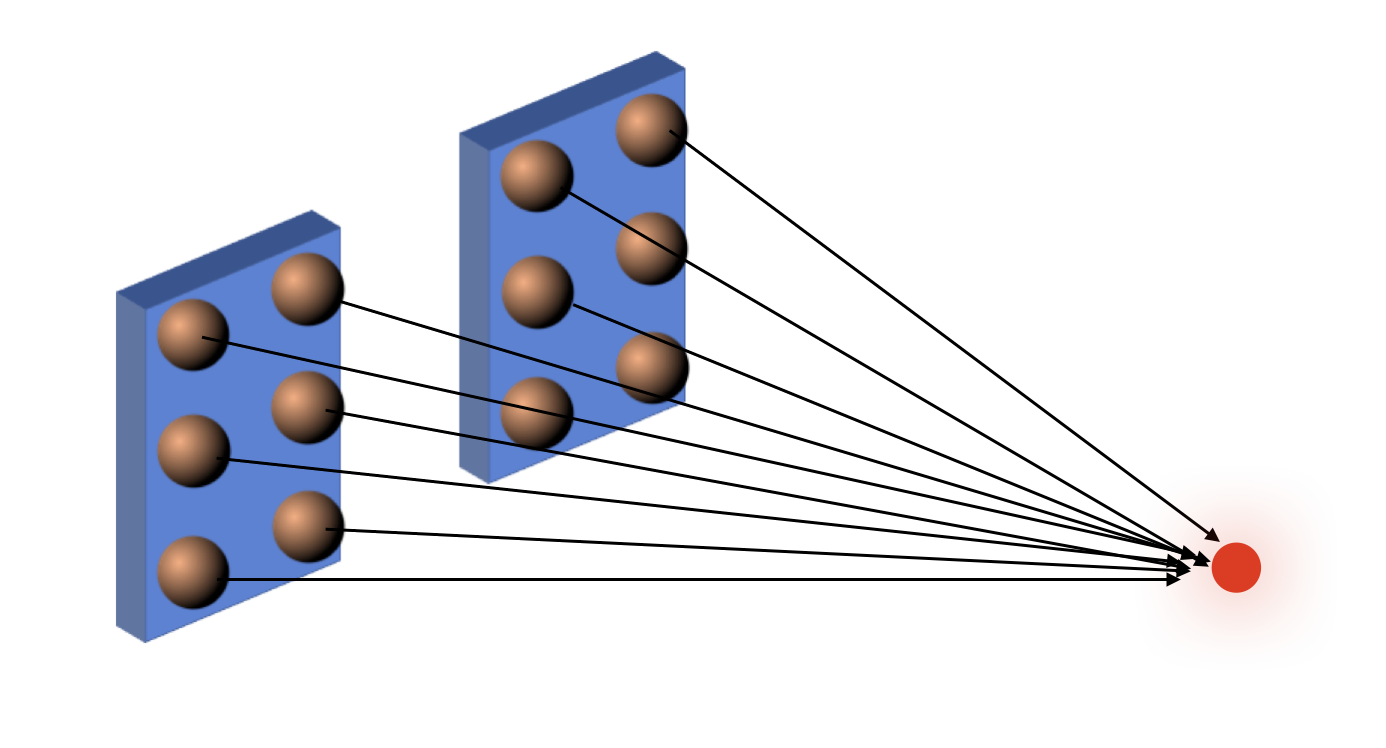}
    \caption{The spherical histograms allow treating the Compton camera as a multi-view camera enabling the reconstruction of the volume.}
    \label{fig:multiview}
\end{figure*}

\subsection{Sparse Projection Operator from Histogram Space to Volume}

Once detector geometry and reconstruction resolution are fixed, each histogram bin becomes a virtual camera element. Let $j=(s,b)$ be the composite index associated with sphere $s$ and angular bin $b$. For each low-resolution bin $j$, a set $\mathcal{M}(j)$ of high-resolution directions is used to define geometric support in the volume. Each support direction is traced through the reconstruction grid and contributes to the sparse system model.

For a support ray $h \in \mathcal{M}(j)$ and a voxel $v$, the contribution can be written as
\begin{equation}
\widetilde{A}_{j,v}^{(h)} = l_{h,v}\,\eta_{s,v}\,\exp(-d_{h,v}/\lambda),
\end{equation}
where $l_{h,v}$ is the voxel-intersection length, $\eta_{s,v}$ is the detector-efficiency weight, and $d_{h,v}$ is the accumulated attenuation distance when attenuation is enabled. The compacted coefficient stored for reconstruction is then
\begin{equation}
A_{j,v} = \sum_{h \in \mathcal{M}(j)} \widetilde{A}_{j,v}^{(h)}.
\end{equation}
The resulting operator is stored in sparse form in both camera-centric and voxel-centric layouts. This avoids on-the-fly cone/voxel traversal during iterative reconstruction and makes forward and backward updates linear in the number of active sparse entries.

\subsection{Snapshot-Based Iterative Reconstruction}

The online acquisition stage continuously updates the spherical histograms, whereas reconstruction operates on a coherent snapshot of those histograms. Let $y_j$ denote the histogram value of bin $j$ in a frozen snapshot and let $x_v$ denote the estimated activity at voxel $v$.

The initialization step computes a normalized weighted backprojection:
\begin{equation}
x_v^{(0)} = \frac{\sum_{j \in \mathcal{N}(v)} A_{j,v}\,y_j}
{\sum_{j \in \mathcal{N}(v)} A_{j,v}},
\end{equation}
where $\mathcal{N}(v)$ is the set of camera bins coupled to voxel $v$. Forward projection at iteration $k$ is
\begin{equation}
\hat{y}_{j}^{(k)} = \sum_v A_{j,v}\,x_v^{(k)},
\end{equation}
and the multiplicative correction can be expressed as
\begin{equation}
x_v^{(k+1)} = x_v^{(k)} \sum_{j \in \mathcal{N}(v)}
\left(
\frac{y_j}{\max(\hat{y}_{j}^{(k)},\epsilon)}
\right)
\frac{A_{j,v}}{\sum_{j' \in \mathcal{N}(v)} A_{j',v}}.
\end{equation}
In practice, the implementation also supports boundary damping, selective camera subsets, and cutoff terms derived from running camera statistics. The main point is that the event stream has already been encoded into histogram space before this iterative stage begins.

\subsection{Low-Latency Execution Model and Complexity}

The proposed pipeline has two computational phases. The first one is online event encoding, where each new Compton event updates the histogram state. The second one is reconstruction from a coherent snapshot through the sparse operator described above. This leads to the cost decomposition
\begin{equation}
T_{\text{ours}} \approx N \cdot C_{\text{cone}\rightarrow\text{sphere}} + T_{\text{setup}} + I \cdot C_{\text{FB}}(E),
\end{equation}
where $N$ is the number of accumulated events, $T_{\text{setup}}$ is the one-time sparse-operator construction cost, $I$ is the number of reconstruction iterations, and $E$ is the active sparse support size. After setup, the dominant iterative cost is governed by $E$ rather than by $N$. This is the key algorithmic distinction with respect to event-by-event list-mode reconstruction, whose dominant cost typically grows with the event count because cone-dependent weights remain inside the inversion loop.

This complexity split is also the one evaluated experimentally in the Results section. The conventional 15-iteration list-mode MLEM reference, implemented internally and still under evaluation as a software implementation, provides the traditional event-driven reconstruction path against which the proposed representational change is compared. The proposed method is instead timed as a snapshot-to-volume transaction after histogram accumulation and sparse-operator setup. The two measurements therefore probe different computational stages and should be interpreted as evidence of different scaling laws, not as a kernel-level apples-to-apples microbenchmark.

The two main execution benefits arise from different parts of the design. Parallelization comes from the object decomposition introduced above: each sphere-local state can be served by its own queue and updated independently. Latency reduction, in contrast, comes from the cone-update algorithm itself, because it reduces the per-event cost required to map a Compton event onto the spherical state. In practical terms, the architecture increases concurrency across queues, while the cone-projection algorithm lowers the service time within each queue.

The decoupling between acquisition and inversion also provides a practical execution model. Histogram states can be frozen and reconstructed asynchronously without replaying the full event list. In other words, the method is not only an accelerator for static reconstruction; it also enables bounded-latency online updates under continuous acquisition.

\subsection{Evaluation Protocol with Monte Carlo Simulation\label{ssec::experiments}}

The evaluation centers on the foundational contribution of the paper: the representational and algorithmic change from list-mode inversion to stateful spherical-histogram reconstruction. Detector-specific refinements such as crystal distortion correction, online self-calibration, and further quality tuning are treated as composable extensions of the same architecture.

We define the \emph{near-field} regime as source distances below twice the diagonal of the combined active detector face ($d < 2D$, where $D$ is the detector diagonal), and the \emph{far-field} regime as distances beyond that threshold ($d \geq 2D$). For the bilateral geometry used in this work, the combined active face spans $92 \times 52$\,mm, giving $D \approx 106$\,mm and a near/far-field boundary of $2D \approx 211$\,mm. The Monte Carlo simulations are carried out using the Geant4 simulation toolkit~\cite{Geant4_2003, Geant4_2006, Geant4_2016} in the near-field regime, where the fly-eye approximation is more demanding. This regime is the most relevant one for the bounded-latency volumetric reconstruction problem targeted in this work.

\subsubsection{Detector Geometry}
The near-field Monte Carlo localization study uses a separate simplified four-block GAGG model, arranged as two bilateral scatter--absorber modules; it is not a simulation of the six-crystal PETsys acquisition chain described above. Each module contains a scatter block of $26 \times 52 \times 5$\,mm and an absorber block of $26 \times 52 \times 8$\,mm, separated by approximately 24\,mm of free space. The two modules are positioned symmetrically at $x = \pm 33$\,mm, forming an open bilateral geometry. The total active scatter area is $2 \times (26 \times 52) = 2{,}704$\,mm$^2$ and the total active absorber area is equal. For the near-field evaluation reported here, a single Cs-137 point source at $662$\,keV is placed at position $(0, 10, 40)$\,mm with respect to the detector reference frame. Compton event reconstruction uses the energy deposits recorded in each scatter--absorber pair to compute the interaction positions and the Compton angle.

The spherical histograms maintain full $4\pi$ angular coverage per sphere. In this bilateral open geometry, however, the useful signal concentrates in the forward hemisphere directed toward the source, and it is this hemisphere that drives the reconstruction. The complementary hemisphere remains available in the encoded state but carries no signal in this configuration.

\subsubsection{Noise Model}
The Monte Carlo event stream is passed through the detector-response convolutor before reconstruction. In addition to the native Geant4 material physics of GAGG at $662$\,keV, this model can introduce detector-specific electronic uncertainty (energy resolution $\sigma_E / E = 8\%$, lateral position uncertainty $\sigma_{x,y} = 2$\,mm, and axial position uncertainty $\sigma_z = 3$\,mm). The relevant observation in the present study is qualitative: introducing this convolution preserves the same convergence behaviour and visible source structure at the chosen reconstruction scale. This does not imply that the electronic uncertainty has no physical effect. Rather, with 4-mm voxels and fly-eye elements covering approximately $4 \times 4$ physical detector pixels, the effective spatial-angular resolution is dominated by the selected voxel, pixel, and sphere separations. At these dimensions, the projection effects discussed here are not resolved more finely than the uncertainty introduced by the electronic-response model; removing the convolution therefore does not visibly improve the reconstruction, whereas including it does not visibly disrupt convergence. A finer spatial sampling or a different detector geometry could make this response contribution observable. The reported localization results should therefore be interpreted in the context of this specific detector and fly-eye configuration. No human body phantom, tissue attenuation, or in-medium scatter is included in the quantitative localization simulations; those experiments are performed in free-field geometry to isolate the behaviour of the reconstruction method itself.

For this paper, the evaluation is organized around three observables:
\begin{itemize}
    \item point-source localization accuracy and, when available, source-separation behaviour in near-field;
    \item bounded-latency reconstruction from coherent histogram snapshots during continuous acquisition;
    \item invariance of reconstruction latency with respect to the number of accumulated events once setup is fixed.
\end{itemize}

Runtime measurements were obtained from an instrumented integrated reference pipeline in which asynchronous histogram accumulation, coherent snapshot synchronization, and continuous volumetric reconstruction operate together under realistic acquisition conditions. In other words, the reported latency values come from the full stateful reconstruction workflow rather than from an artificially isolated kernel benchmark. This choice is methodological rather than conceptual: the integrated pipeline is used here as measurement infrastructure for input event rate, snapshot-to-volume transaction cost, and sustained update cadence, but it is not itself presented as a separate scientific contribution. The distinction between lower-rate and higher-rate acquisition regimes is therefore operational rather than algorithmic. Unlike list-mode reconstruction, the proposed method does not place an ever-growing event history back inside the inversion loop: incoming events are absorbed into a persistent histogram state, and reconstruction operates on coherent snapshots of that state. As a consequence, changing acquisition rate modifies the external operating conditions of the pipeline, but not the structure of the reconstruction algorithm itself.

The acquisition rates of the sessions reported here are set by the measurement conditions of these particular experiments and are well below the input rate that the online encoding stage can sustain; they should therefore be read as operating points of the reported sessions, not as a throughput limit of the method. In the encoding stage the per-event cost is bounded by the discrete circular support of the cone on the sphere, so the acquisition side is not the limiting factor in the regimes considered here.

Monte Carlo generation can be parallelized on a computing cluster to deliver event streams well above the online needs of the reconstructor. Reported performance values, however, must refer to the hardware on which the reconstruction pipeline itself is executed. Real-time videos and executable applications are therefore treated as supplementary operational evidence of bounded-latency behaviour, not as substitutes for quantitative localization or timing metrics.

\begin{table}[t]
\centering
\caption{Consolidated evaluation configuration. The phantom and Monte Carlo experiments serve different purposes and use distinct detector descriptions.}
\label{tab:evaluation_configuration}
\footnotesize
\begin{tabular}{p{0.34\columnwidth}p{0.56\columnwidth}}
\hline
Component & Configuration \\
\hline
Measured phantom & $^{18}$F-FDG, 511-keV photons; 5-min acquisition; $8\times8\times8$~cm$^3$ box with approximately 1.9~MBq total activity. \\
Physical detector & Six GAGG modules: three $25.5\times51.0\times5.0$~mm$^3$ scatter and three $25.5\times51.0\times8.0$~mm$^3$ absorber modules; PETsys readout. \\
Event formation & Pixel and event grouping windows: 10,000~ps each; 4\% energy selection at 511~keV. \\
Near-field Monte Carlo & Geant4, $^{137}$Cs at 662~keV; separate simplified four-block GAGG geometry; free-field conditions. \\
CES reconstruction & 4-mm voxels; maximum 20 MLEM iterations; operational stop when the successive-update metric is below $10^{-4}$ (typically 10--15 iterations in the reported sessions). \\
Conventional list-mode MLEM reference & 15 fixed iterations; internal implementation under evaluation; NVIDIA GeForce RTX~2080~SUPER GPU on an AMD Ryzen~9~3900X host. \\
Phantom bias reduction & Relative-bias subtraction with $\tau=0$, $0.7$, and $0.8$; qualitative stability demonstration only. \\
\hline
\end{tabular}
\end{table}

\section{Results}

\subsection{Near-Field Reconstruction}

Near-field reconstruction is the primary validation case of this work because it is the geometrically more demanding regime for the fly-eye approximation and the most computationally ambitious one for volumetric localization. The main quantitative observable in this regime is centroid-based point-source localization error measured from saved reconstructed volumes against Monte Carlo source positions rasterized on the same reconstruction grid. Peak-based localization error is retained as a secondary control metric.

Four independent snapshots were analysed across four acquisition sessions, all with the point source at $(0,10,40)$\,mm (Table~\ref{tab:nearfield_localization}). Three of the four snapshots show a consistent centroid-based localization error in the range 3.5--3.8\,mm. In the fourth snapshot (session bddef1c6, 74\,528 events, layer~1), the peak is well-localized at $(-2,10,40.6)$\,mm with a peak error of 2.1\,mm, but the centroid is pulled to a secondary diffuse lobe, yielding a centroid error of 16.0\,mm. This case illustrates that the centroid metric is susceptible to background lobes at lower-resolution reconstruction layers, while the peak metric remains robust. In the three consistent snapshots, centroid and peak errors are complementary: centroid error spans 3.5--3.8\,mm and peak error spans 4.5--6.4\,mm. No thresholding or dedicated volumetric post-processing was applied; volumes are analysed directly in the state produced by the iterative solver.

\begin{table}[t]
\centering
\caption{Near-Field Point-Source Localization Results. Source position: $(0,10,40)$\,mm. All snapshots used 20 MLEM iterations on a $33\times33\times29$ voxel grid (4\,mm spacing). Centroid computed over voxels above 50\% of peak value.}
\label{tab:nearfield_localization}
\begin{tabular}{lrrlrr}
\hline
Session & Events & Layer & Centroid (mm) & Peak err. & Centroid err. \\
\hline
7c9fa110 & 100\,012 & 2 & $(-2.23,\,8.23,\,42.37)$ & 4.5\,mm & 3.7\,mm \\
6ac4cb5d & 100\,025 & 1 & $(-2.00,\,8.36,\,42.37)$ & 6.4\,mm & 3.5\,mm \\
f53e7dcd &  24\,192 & 1 & $(-2.26,\,8.28,\,42.50)$ & 6.4\,mm & 3.8\,mm \\
bddef1c6 &  74\,528 & 1 & $(-2.10,\,16.84,\,54.32)$ & 2.1\,mm & 16.0\,mm$^*$ \\
\hline
\multicolumn{6}{l}{$^*$ Secondary diffuse lobe pulls centroid away from hotspot.}
\end{tabular}
\end{table}

The consistent centroid errors of 3.5--3.8\,mm across three independent snapshots with event counts ranging from 24\,192 to 100\,025 support the interpretation that the representational method recovers the global source location reliably at this near-field distance and reconstruction configuration. The outlier case reinforces the joint use of both metrics: when peak and centroid diverge significantly, a secondary lobe is present and the peak provides a more reliable localization estimate.

\subsection{Low-Latency Online Reconstruction and Event-Count Invariance}

The main performance result is not a single wall-clock number detached from the acquisition mode, but the existence of a bounded reconstruction latency once the histogram state and sparse operator are in place. To quantify this behaviour, we ran a reference continuous session of 281.9\,s (session da19e991, 2026-04-24) in which acquisition and reconstruction remained active in parallel. During this session, 247,830 raw events were acquired. The reconstruction pipeline produced 103 coherent snapshots, each with a maximum budget of 20 MLEM iterations and the operational stopping rule described in Table~\ref{tab:evaluation_configuration}; the observed runs typically used 10--15 iterations. The mean transaction cost was $2{,}012 \pm 82$\,ms (CV~4.1\%) on an AMD Ryzen~9 3900X workstation with 128\,GiB RAM and no GPU acceleration.

Crucially, the reconstruction cost remained in the same range throughout the session even as the accumulated acquisition history grew by several orders of magnitude. The first snapshot was reconstructed after only 55 raw events; the last after 243,543 raw events and 12,536,320 accumulated histogram-space events --- a factor of about $4.4\times10^{3}$ in raw events and $2.3\times10^{5}$ in accumulated histogram-space contributions, with no comparable increase in reconstruction cost. The first five snapshots averaged 1,977\,ms, the full set of 103 averaged 2,012\,ms, and the last five averaged 2,105\,ms (Table~\ref{tab:latency_invariance}).

\begin{table}[t]
\centering
\caption{Reconstruction Transaction Cost Across the Reference Session (da19e991, 2026-04-24, 247,830 events). Snapshots used a maximum of 20 MLEM iterations with the operational stopping rule in Table~\ref{tab:evaluation_configuration}; no GPU acceleration applied.}
\label{tab:latency_invariance}
\begin{tabular}{lrrr}
\hline
Stage & Snapshots & Mean (ms) & Std.\ dev.\ (ms) \\
\hline
First 5 snapshots & 5   & 1977 & 121 \\
All snapshots     & 103 & 2012 &  82 \\
Last 5 snapshots  & 5   & 2105 &  24 \\
\hline
\end{tabular}
\end{table}

\subsection{Measured Linear Event-Count Dependence of Conventional List-Mode Reconstruction}

To compare the proposed representation with the conventional reconstruction paradigm, we measured a dedicated GPU benchmark of our internally implemented list-mode MLEM reference. Its software implementation remains under evaluation, but it represents the traditional event-driven reconstruction path relevant to the central comparison of this paper. The same calibration input was used across 120 runs grouped into 12 event-count levels from $10^3$ to $80{,}701$ events. Each reconstruction used 15 MLEM iterations. The mean runtime increases monotonically from 7.053\,s at 1,000 events to 372.121\,s at 80,701 events (Table~\ref{tab:listmode_linear}). A least-squares fit over all 120 runs gives
\begin{equation}
T_{\text{LM}}(N) \approx 2.585 + 4.580\times 10^{-3} N \quad \text{s},
\end{equation}
with $R^2 = 0.9999987$, which is effectively linear over the measured range.

\begin{figure*}[t]
    \centering
    \includegraphics[width=\textwidth]{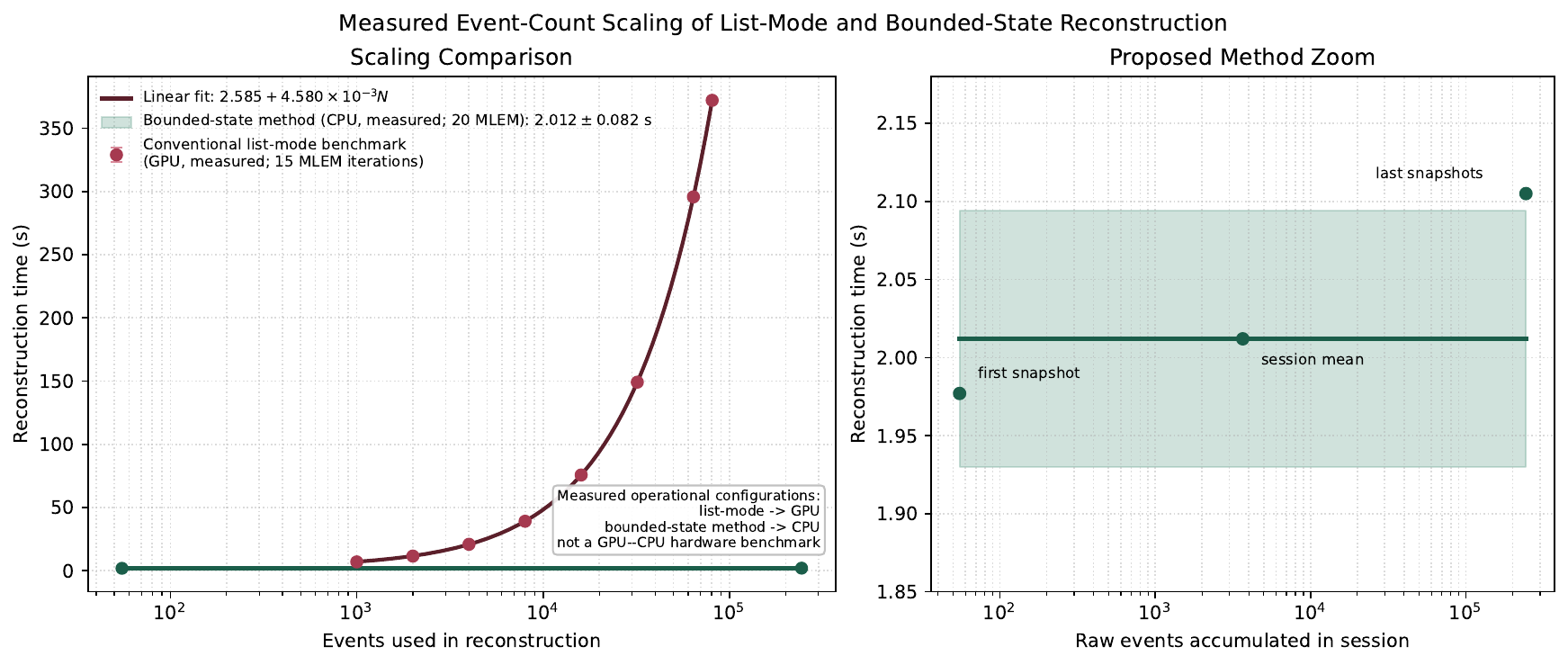}
    \caption{Measured event-count scaling in the two operational reconstruction configurations.
    \textbf{Left:} reconstruction time as a function of the number of events used in reconstruction.
    GPU list-mode timings (red circles, 15 MLEM iterations, $R^2 = 0.9999987$)
    follow the fitted linear trend $T_{\text{LM}} \approx 2.585 + 4.580\times10^{-3}N$\,s. The proposed
    bounded-state method (green band, maximum 20 MLEM iterations with operational stopping, CPU-only) stays at $2.012 \pm 0.082$\,s across the full range
    of accumulated events. \textbf{Right:} zoom on the proposed method showing the first snapshot
    (55\,events), the session mean, and the last snapshot (243,543\,events). The near-constant
    reconstruction cost despite this growth of the accumulated state directly demonstrates
    event-count invariance of the snapshot-to-volume transaction. The figure does not constitute a controlled
    comparison of GPU and CPU hardware efficiency.}
    \label{fig:scaling_comparison}
\end{figure*}

This result establishes the central architectural contrast of the paper: a full list-mode inversion keeps its wall-clock cost tied to the number of events carried into reconstruction, whereas the histogram-snapshot measurements in Tables~\ref{tab:latency_invariance} and~\ref{tab:rate_invariance} remain near 2\,s after events have been absorbed into the bounded spherical state. Thus, event count primarily affects the online accumulation stage, while iterative reconstruction from the encoded state is controlled by the sparse operator and iteration budget (Fig.~\ref{fig:scaling_comparison}).

The GPU list-mode and CPU bounded-state measurements are reported to expose this difference in event-count dependence, not as a hardware-performance benchmark or a fully harmonized comparison of absolute throughput. The present paper introduces the bounded spatial-angular representation; a dedicated study with matched implementations, stopping policies, image-quality targets, and accelerator configurations would be required to compare hardware efficiency independently of that representational change.

\begin{table}[t]
\centering
\caption{Dedicated GPU Timing Benchmark of the Conventional List-Mode MLEM Reference (15 iterations). Representative event-count levels are listed; each level was repeated 10 times on the same calibration input, using an NVIDIA GeForce RTX~2080~SUPER GPU hosted by an AMD Ryzen~9~3900X 12-core workstation. The internal software implementation remains under evaluation. This measurement supplies the conventional event-driven reference for the central scaling comparison; it is not a controlled GPU--CPU hardware benchmark.}
\label{tab:listmode_linear}
\begin{tabular}{rrr}
\hline
Events & Mean time (s) & Std.\ dev.\ (s) \\
\hline
1,000  & 7.053   & 0.034 \\
2,000  & 11.649  & 0.048 \\
4,000  & 20.965  & 0.025 \\
8,000  & 39.203  & 0.042 \\
16,000 & 75.802  & 0.040 \\
32,000 & 149.214 & 0.172 \\
64,000 & 295.759 & 0.098 \\
80,701 & 372.121 & 0.329 \\
\hline
\end{tabular}
\end{table}

\subsection{Reconstruction Cost is Independent of Acquisition Rate}

To verify that reconstruction cost is decoupled from acquisition throughput, the measurement was repeated across nine independent sessions conducted on three separate dates with varying acquisition rates (Table~\ref{tab:rate_invariance}). In total, 305 snapshots were collected. The mean reconstruction cost across all sessions was $2{,}003 \pm 113$\,ms (CV~5.6\%), consistent across dates despite differences in acquisition rate and accumulated event count. This confirms that the bounded-latency property is reproducible and not an artefact of a single favourable run.

\begin{table}[t]
\centering
\caption{Reconstruction Transaction Cost Grouped by Date Across Nine Independent Sessions.
All sessions used a maximum of 20 MLEM iterations with spatial-subset rotation and the operational stopping rule in Table~\ref{tab:evaluation_configuration}; no GPU acceleration.
Hardware: AMD Ryzen~9 3900X, 128\,GiB RAM, Ubuntu 22.04 LTS.}
\label{tab:rate_invariance}
\begin{tabular}{lrrrr}
\hline
Date & Sessions & Snapshots & Mean cost (ms) & CV (\%) \\
\hline
2026-04-24 & 2 & 113 & $1{,}999 \pm 84$  & 4.2 \\
2026-04-29 & 2 &  45 & $1{,}933 \pm 73$  & 3.8 \\
2026-04-30 & 5 & 147 & $2{,}026 \pm 131$ & 6.4 \\
\hline
All combined & 9 & 305 & $2{,}003 \pm 113$ & 5.6 \\
\hline
\end{tabular}
\end{table}

This computational decoupling is relevant to prospective medical-physics applications of Compton cameras, including online dose monitoring, prompt-gamma range verification, and radiotracer-guided surgery. In such applications, a higher usable detector count rate could support shorter acquisition windows or higher temporal sampling, subject to detector performance, image-quality requirements, and the clinical measurement environment. With full list-mode reconstruction, a higher count rate enlarges the event set carried into each inversion and can make sustained online operation increasingly costly. In the bounded-state representation, incoming events are instead encoded online and the reconstruction stage operates on a fixed-dimensional snapshot.

The results in Table~\ref{tab:rate_invariance} directly demonstrate this property for the tested detector configuration and acquisition-rate range: across sessions acquired on different dates and with different input rates, the snapshot-to-volume latency remains essentially unchanged. They do not constitute a clinical validation or establish patient-dose, image-quality, or tissue-attenuation performance. Rather, they show that, provided the detector and online encoding remain within their valid operating regimes, increasing the usable event rate need not increase the latency of the bounded-state reconstruction stage. Validation of this potential benefit with patient-equivalent attenuation, scatter, and clinically relevant count-rate conditions remains future work.

\subsection{Structured Phantom Reconstructions}

The structured phantom experiment presented here serves a specific purpose within a framework paper: to establish that the \ac{CES} representation remains operational for spatially distributed activity patterns, and that it does so in a case in which the conventional list-mode Compton reconstruction pipeline previously applied to the same detector and phantom configuration did not recover separable source structure. The scientific interest of the result is therefore not the absolute image quality achieved, but the stability of coherent volumetric structure under deliberately simple processing: a basic iterative solver followed by a relative-bias background-subtraction stage, without advanced regularisation, learned denoising, or phantom-specific tuning. The bias subtraction is used here as a minimal diagnostic reduction, not as the noise-treatment contribution of the paper. New reconstruction methods are being developed separately; they are not proposed or evaluated here and will be reported in a separate study only after validation.

The phantom is intentionally non-trivial. It consists of three spherical liquid sources filled with residual FDG solution from a PET procedure and immersed in a diluted-FDG background contained in an $8 \times 8 \times 8$~cm$^3$ box. The total activity in this box was approximately 1.9~MBq at acquisition. The nominal concentration ratio between the spheres and the background was 4, while the successive dilutions make the activity of each individual sphere unavailable as a traceable quantitative value. The phantom is therefore used as a qualitative structured-activity test rather than as an activity-quantification experiment. The primary sequence is reconstructed without attenuation correction, and a second exploratory sequence applies a simple water-equivalent attenuation compensation in the sparse projection operator. This uniform-water model is deliberately simple, but it is also physically relevant as a first-order approximation to soft-tissue and fluid environments in medical imaging. Residual source displacement and diffuse background structure are therefore expected. The case combines multiple sources, non-uniform activity concentration, overlapping volumetric support, distributed background, and imperfect attenuation modelling simultaneously --- conditions under which the conventional pipeline did not produce separable structure.

Reconstructed volumes were processed by a simple relative-bias background subtraction,
\begin{equation}
x_\tau(v)=\max\left(x(v)-\tau\max_v x(v),0\right),
\end{equation}
where $x(v)$ is the reconstructed activity at voxel $v$ and $\tau$ defines a bias level relative to the maximum reconstructed value. The operation estimates the background bias as $\tau\max_v x(v)$, subtracts it from every voxel, and clips negative values to zero. The case $\tau=0$ applies no bias subtraction; increasing $\tau$ removes a larger background component. It also clips and reduces the amplitude of the retained source signal, as follows directly from the subtraction in the equation; it therefore does not preserve quantitative activity. This deterministic background-removal algorithm is deliberately simple: it tests whether the reconstruction residual in \ac{CES} space can be reduced while recognizable coherent source structures remain.

\begin{figure}[t]
    \centering
    \includegraphics[width=0.95\columnwidth,clip]{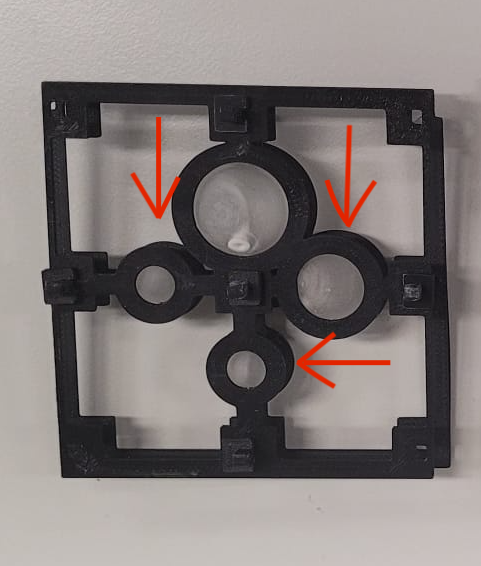}
    \caption{Physical three-sphere phantom configuration. Sources are spherical 
    liquid containers filled with residual FDG solution from a PET procedure and immersed 
    in a diluted-FDG background in an $8 \times 8 \times 8$~cm$^3$ box (total activity approximately 1.9~MBq at acquisition). The nominal sphere-to-background concentration ratio is 4; individual source activities are not used as quantitative ground truth. The qualitative reconstruction sequence 
    shown in Fig.~\ref{fig:phantom_thresholds} corresponds to this phantom.}
    \label{fig:physical_phantom}
\end{figure}

\begin{figure*}[t]
    \centering
    \includegraphics[width=\textwidth]{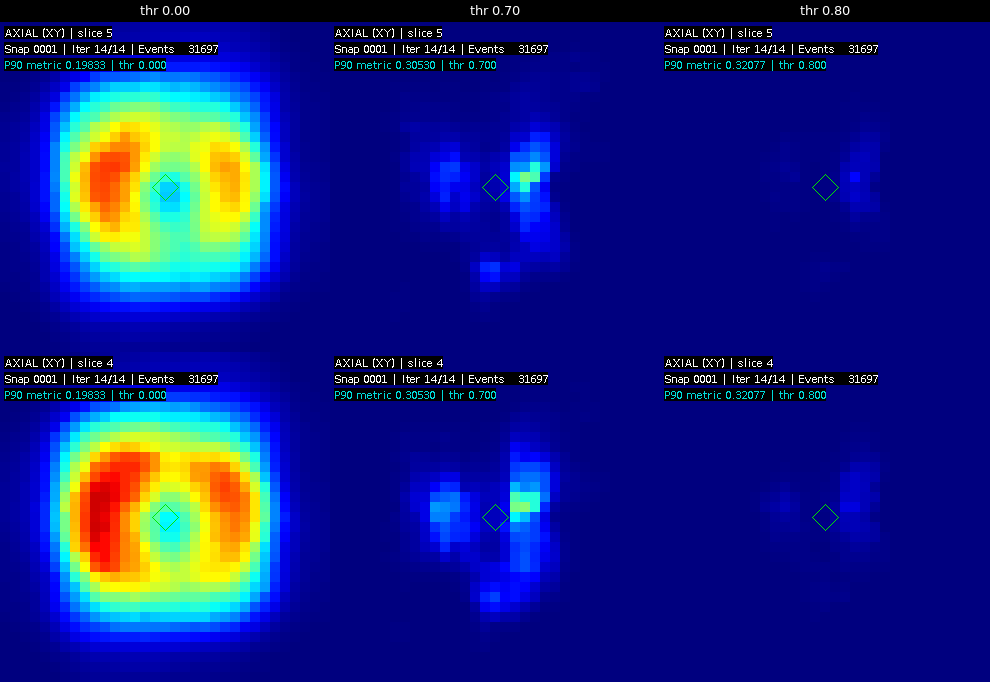}
    \caption{\ac{CES} structured-phantom reconstruction after relative-bias background subtraction (axial view, iteration~14). Columns correspond to $\tau=0$, $0.7$, and $0.8$; top row: no attenuation correction (\emph{fantoma\_serie\_images\_000}); bottom row: water-equivalent attenuation compensation applied to the sparse projection operator ($\mu_\text{water}=0.096\,\text{cm}^{-1}$, mean free path $\approx104$\,mm at $511\,\text{keV}$, \emph{fantoma\_serie\_images\_007}). The $\tau=0$ column shows the uncorrected cone-shadow background. It is a \ac{CES} result, qualitatively compared in the text with the separately supplied list-mode sequence, rather than a side-by-side list-mode reconstruction. At $\tau=0.7$, subtracting the relative bias reveals the sphere arrangement; at $\tau=0.8$, further subtraction also reduces source contrast. The panels are qualitative examples of stability under this minimal reduction, not a systematic noise or denoising study.}
    \label{fig:phantom_thresholds}
\end{figure*}

The qualitative sequence shown here corresponds to the reconstructed phantom series \emph{fantoma\_serie\_images\_000} for the physical phantom shown in Fig.~\ref{fig:physical_phantom}. At $\tau=0$, the initial \ac{CES} volumetric projection is dominated by cone-shadow-like residual contributions: their accumulated weight produces a high-intensity feature on the side opposite the physical source arrangement. This uncorrected \ac{CES} behaviour corresponds qualitatively to that of the conventional reconstruction. The $\tau=0$ column of Fig.~\ref{fig:phantom_thresholds} displays the uncorrected \ac{CES} result, not a list-mode panel; the conventional sequence is provided separately in the supplementary reconstruction videos. In that list-mode sequence, subsequent iterations did not displace the dominant feature towards the sphere locations, and separable source structure was not recovered.

The same initial \ac{CES} projection contains a different separation of signal and residual. Applying the simple $\tau=0.7$ relative-bias subtraction removes a substantial part of the cone-shadow background, while also cutting the retained signal amplitude; nevertheless, the first background-reduced projection already reveals the three-sphere arrangement. Subsequent iterations then converge toward the displayed source-like structures. The $\tau=0.7$--$0.8$ views in Fig.~\ref{fig:phantom_thresholds} therefore demonstrate a qualitative distinction in the residual distribution: recognizable coherent source-like components persist, whereas much of the diffuse cone residual is removed as bias. The algorithm is deliberately simple and is not used for activity quantification; its role here is to demonstrate stability of the encoded representation under a minimal background reduction, rather than to establish a final denoising solution. More complex methods are separate work and require their own validation.

The attenuation-compensated sequence adds a useful stress test. In cone-based reconstruction spaces, attenuation weighting can increase not only physically useful source contributions but also inconsistent cones whose support is brought closer to the source region, thereby increasing apparent background or structured noise. In the \ac{CES} reconstruction, the same compensation improves the contrast and spatial arrangement of the surviving structures once the relative threshold is applied. The relevant observation is therefore not that attenuation compensation alone solves the phantom, but that attenuation weighting and spatial-angular consistency filtering become complementary in the encoded representation. This complementarity is especially important for medical applications, where attenuation, distributed background activity, and heterogeneous tissue paths are intrinsic features of the imaging problem rather than secondary corrections.

A plausible reason why the \ac{CES} representation enables this behaviour, while conventional list-mode reconstruction does not in this experiment, lies in the geometry of the background contribution. In classical Compton reconstruction, each event contributes a full cone backprojected into the volume. The integrated cone background --- the aggregate of all back-projected cones from background events --- distributes diffuse activity throughout the reconstruction volume with limited locality. This diffuse cone background is a major obstacle to interpretable reconstruction under high-background conditions: it grows as events are added and is difficult to separate from source signal without explicit modelling of the contributing event population.

In the \ac{CES} representation the same event contributes a circle on a local geodesic sphere before volumetric inversion. Because each sphere encodes angular directions from a specific detector position, diffuse background and true localized sources populate the encoded state with different multi-view consistency. The volume-level relative-bias subtraction therefore probes that encoded structure: source-like components that receive coherent support across several sphere-local angular domains tend to remain recognizable after background removal, whereas diffuse cone residuals are preferentially removed. This is a simple background-removal algorithm, evaluated here qualitatively rather than through an exhaustive image-quality study. Its behaviour is consistent with the interpretation of each sphere as a local angular coherence gate whose combined action across the fly-eye arrangement makes incoherent background more separable after reconstruction. The water-equivalent attenuation compensation further modifies the sparse system weights and improves the background-reduced visual separation in this example, although the uniform-medium assumption remains incomplete for a heterogeneous physical phantom. The comparison also illustrates why attenuation modelling and coherence filtering should not be treated as independent corrections: attenuation can raise the amplitude of both useful and inconsistent contributions, while \ac{CES} coherence provides a mechanism to preferentially retain the former.

Reaching clinical image quality for this class of phantom --- with proper attenuation correction, scatter modelling, advanced regularisation, and validated denoising --- is a programme of work that builds on the representational architecture presented here, not a claim of the present paper. What this experiment establishes is a more foundational point: the \ac{CES} representation reorganizes the inverse problem into a state where meaningful volumetric structures remain spatially and angularly correlated, while a substantial part of the incoherent residual becomes separable by even the simplest consistency criterion. This property was not accessible to the conventional list-mode pipeline under the same experimental conditions, and it is a direct consequence of the encoded spatial-angular structure of the \ac{CES} state.

This observation connects directly with the broader interpretation of \ac{CES} as a computational hologram. Because the encoded state distributes source information across many detector-centred angular observations, coherent sources produce correlated contributions across the representation, whereas diffuse or incoherent components do not. This provides a principled starting point for separate, subsequently validated studies of multi-view coherence filtering, physically constrained denoising, or ML-based reconstruction operating on the bounded tensor-compatible \ac{CES} state.

\section{Discussion}

The main contribution of this work is a change in computational representation. Conventional list-mode reconstruction keeps the event stream in its raw form and therefore carries event-dependent physics into the inversion stage. Our approach encodes each event first into a detector-centred spherical state and reconstructs later from that state through a sparse operator. This means that the relevant computational object is no longer the full event list, but a bounded spatial-angular representation plus a calibrated inverse model in memory. In conceptual terms, this bounded state behaves as a computational hologram: a coded angular surface that retains sufficient information for volumetric inversion while avoiding repeated event-wise traversal during reconstruction.

The hologram terminology is intended as an architectural analogy rather than as a statement about optical interference or coherent phase reconstruction. The volumetric information is distributed across many correlated sphere-local angular observations, each observed from a different detector position and therefore with different parallax. Reconstruction recovers the activity volume from the consistency of these distributed observations. Equivalently, the \ac{CES} may be viewed as a sparse localized frame over a detector-position and angular-direction product space, closer in spirit to a distributed directional field than to a global spherical-harmonic expansion.

This distinction is important because it changes scalability. The dominant cost of reconstruction is shifted from repeated cone/voxel processing to sparse forward/backward passes over the active model. The method is therefore a route toward faster and more predictable Compton reconstruction under continuous acquisition.

Precomputation of a system matrix is standard in tomographic modalities with discrete, enumerable measurement geometries, such as PET. In Compton cameras, however, each detected event defines a cone with a continuously varying axis and opening angle, making the list-mode system model inherently event-dependent. As a result, a conventional list-mode reconstruction input continues to grow with the number of retained events even when parts of the geometry or response model are precomputed. The spherical-histogram encoding introduces the missing fixed discrete intermediate space: angular bins attached to detector-centred spheres. Once events have been accumulated into this bounded \ac{CES} snapshot, a static sparse operator from histogram space to volume can be defined and precomputed. This structural change is what enables bounded-latency reconstruction.

This point can be contrasted in a controlled way with a fully matched conventional list-mode MLEM workflow, because the two approaches differ not only in implementation cost but in computational scaling law. The present article provides the key scaling evidence: the dedicated 15-iteration list-mode benchmark is linear in event count over the tested range, whereas the proposed method exhibits near-invariant snapshot reconstruction time after state formation and setup. A tighter one-to-one benchmarking protocol with fully harmonized stopping policies, detector subsets, and implementation details would then address absolute throughput comparison.

The GPU aspect follows naturally from the same architectural separation. The current results were obtained without GPU acceleration, yet the proposed formulation is structurally more amenable to GPU execution than a cone list-mode inversion: the encoded histogram state, the sparse inverse operator, and the iterative update kernel are fixed, bounded computational objects. After state formation, the dominant work consists of repeatable sparse forward/backward passes and regular volumetric updates over bounded arrays, with reusable data layout and without carrying a growing event list through each iteration. By contrast, GPU-accelerated list-mode MLEM retains event-dependent cone processing and memory traffic that grow with the list preserved inside the inversion loop. This is an architectural GPU-friendliness claim, not a measured GPU speedup for the present implementation; a dedicated GPU implementation and matched benchmark would quantify its practical magnitude.

The same representation has consequences beyond the present MLEM implementation. Because \ac{CES} snapshots are fixed-dimensional, detector-centred, sparse, and geometrically interpretable, they are naturally compatible with tensor-based processing and differentiable reconstruction pipelines. This suggests a path toward denoising methods that operate on multi-view spatial-angular coherence rather than on isolated list-mode events. In that setting, physically consistent signal would be reinforced by correlations across spheres, while incoherent components could be suppressed before or during volumetric inversion. These extensions follow directly from the structure of the encoded state.

The attenuation-compensated phantom result is relevant in the same sense. Medical Compton imaging cannot rely on free-field assumptions: attenuation, scatter, organ geometry, and activity background are part of the forward problem. Because many soft-tissue and lesion environments are water-like to first order at $511\,\text{keV}$, a water-equivalent correction is a meaningful initial model rather than an arbitrary numerical adjustment. The preliminary phantom sequence suggests that \ac{CES} provides a natural place to combine attenuation-aware system weights with spatial-angular coherence criteria, so that compensation of physically plausible signal does not simply amplify inconsistent cone background. A clinically validated attenuation and scatter model remains future work, but the architectural requirement is already visible in the present experiment.

Near-field validation is central in this paper because it is the more demanding regime for both geometry and computation in the intended bounded-latency reconstruction setting. This allows the paper to make a focused claim without trying to exhaust all detector corrections, calibration effects, or downstream applications in a single manuscript.

As a framework paper, the present article evaluates the technique in a deliberately minimal reconstruction setting. Reported volumetric results are obtained with a basic iterative MLEM workflow and the simple relative-bias background subtraction described above, without advanced regularisation, learned denoising, or dedicated hotspot-sharpening stages. This isolates the scientific question addressed here: whether the spherical-histogram state representation is operational, quantitatively meaningful, and compatible with bounded-latency reconstruction in its minimal form. Accordingly, the paper reports representative quantitative cases that reflect the stable behaviour observed during device development while keeping exhaustive optimisation as a downstream task.

Several extensions naturally build on this foundation, including detector distortion correction, online self-calibration, matched-quality comparisons against external baselines, and broader detector-specific optimization. Likewise, real-time videos and interactive applications should be understood as supplementary operational evidence. The primary scientific evidence of the present paper comes from localization error, latency decomposition, and event-count invariance. The structured-phantom sequence complements these results qualitatively by showing that the same representation remains usable for distributed activity patterns.

Following the convention of framework papers in tomographic reconstruction, we focus here on the representational change and its scaling properties. Advanced image-quality optimization, systematic denoising comparison, GPU implementation, detailed comparisons with OS-MLEM and learning-based methods, and clinical evaluation constitute a separate research programme that builds on the architecture presented here.

The quantitative point-source simulations reported here use a free-field geometry without a human body phantom or tissue attenuation and scattering medium between the source and the detector. The reported localization errors therefore characterize the reconstruction method under controlled geometric conditions rather than patient-specific or tissue-equivalent clinical accuracy. The structured-phantom result is included as qualitative evidence of distributed-source behaviour, not as a clinical validation study. Incorporating realistic body attenuation, Compton scatter in tissue, and organ-specific geometries is the next clinical-validation layer built on top of the present architecture. The goal of this paper is to demonstrate that the representational and algorithmic change --- from list-mode event inversion to stateful spherical-histogram reconstruction --- is operationally sound, quantitatively meaningful, and compatible with bounded-latency online operation.

\section{Conclusion}

This work has introduced spherical-histogram reconstruction as a change of computational representation for Compton cameras, and has evaluated it against the conventional list-mode MLEM path on the two limitations that most restrict that path in practice.

The first limitation is reconstruction time. The dedicated benchmark reported here shows that list-mode inversion remains linear in the number of events carried into the inversion loop ($R^2 = 0.9999987$ over the measured range), whereas reconstruction from a coherent histogram snapshot stays near $2$\,s across 305 snapshots and nine independent sessions, with the accumulated state growing by several orders of magnitude in between. Once geometry, sparse operator, and iteration budget are fixed, the snapshot-to-volume transaction is therefore effectively independent of the number of accumulated events.

The second limitation concerns what the reconstruction is able to recover. In the structured-phantom experiment, the encoded state retained separable three-sphere structure under a deliberately minimal processing chain, in a configuration in which the conventional list-mode pipeline previously applied to the same detector and phantom did not recover separable source structure. This is reported as a qualitative observation rather than as a systematic image-quality study, but it indicates that the reorganization of the inverse problem is not only a matter of speed: coherent source contributions and diffuse cone residual become separable in the encoded spatial-angular state in a way that they are not in the raw event list.

Both properties were observed across the two evaluation regimes used in this paper: Geant4 Monte Carlo point sources at $662$\,keV with $^{137}$Cs, and measured acquisitions with $^{18}$F-FDG at $511$\,keV on a PETsys-instrumented GAGG detector --- that is, with isotopes in routine use rather than only in simulation.

Because \ac{CES} snapshots are bounded, fixed-dimensional, and geometrically interpretable, they also constitute an acquisition state that is directly processable by tensor-based and learning-based methods. Two directions follow naturally and are left to future work: the reduction of the characteristic Compton background produced by the excess of back-projected cones, and the use of machine-learning or AI-assisted operators acting on the encoded state to improve reconstruction quality. Both require validation programmes of their own and exceed the scope of these first studies of a newly introduced representation.

\section*{Supplementary Material}
Reconstructed volumes, real-time videos, and interactive visualizations of the near-field experiments reported in this work are available at: \url{https://medphysics.i-do.science/demo/prostate_txt_source_0_0_100/}
Structured phantom reconstruction demos are available at:
\url{https://medphysics.i-do.science/demo/fantoma_00/},
\url{https://medphysics.i-do.science/demo/fantoma_01/},
\url{https://medphysics.i-do.science/demo/fantoma_03/},
\url{https://medphysics.i-do.science/demo/fantoma_05/},
\url{https://medphysics.i-do.science/demo/fantoma_07/}, and
\url{https://medphysics.i-do.science/demo/fantoma_08/}.

Reconstructions with water-equivalent attenuation compensation ($\mu_\text{water}=0.096\,\text{cm}^{-1}$ at $511\,\text{keV}$) applied to the sparse projection operator are available at:
\url{https://medphysics.i-do.science/demo/fantoma_serie_images_000_attenuation/},
\url{https://medphysics.i-do.science/demo/fantoma_serie_images_007_attenuation/}, and
\url{https://medphysics.i-do.science/demo/fantoma_serie_images_008_attenuation/}.

\section*{Acknowledgment}
This work was partially funded by ENRESA through the agreement for the development of the research project \emph{Imagen gamma: implementaci\'on de nuevos desarrollos e integraci\'on con dispositivos empleados por Enresa}. J.E. is supported by a predoctoral contract funded by Generalitat Valenciana and European Social Fund, and by the Consejo de Seguridad Nuclear (CSN) through the project \emph{Proton: evaluaci\'on tomogr\'afica de residuos nucleares}.

\section*{Conflict of Interest}
F.~J.~Albiol, A.~Albiol, L.~Caballero, and S.~Tortajada are named inventors on Spanish patent application P202430153, \emph{M\'etodo de procesamiento de im\'agenes tridimensionales que permite minimizar distorsiones}, filed with the Spanish Patent and Trademark Office (OEPM) on 1 March 2024, which covers the geodesic-sphere representation used in this work. The application was filed as a service invention and is co-owned by CSIC, Universitat de Val\`encia, Universitat Polit\`ecnica de Val\`encia, and ENRESA; ENRESA also partially funded this work, as stated in the Acknowledgment. The named inventors hold no personal ownership share. The remaining authors --- J.~Escalante, E.~Larrea Estrelles, and J.~L.~Legan\'es-Nieto --- have no ownership interest in the patent and declare no conflict of interest. No author has any other financial or non-financial interest --- including stock ownership, company board membership, advisory board membership, or paid consultancy --- that could be perceived as influencing the objectivity of this work, beyond the patent interest disclosed above.

\bibliographystyle{IEEEtran}
\bibliography{references}

\end{document}